\documentclass{elsart}
\usepackage{natbib,epsf}
\begin{document}
\runauthor{Atoyan and Dermer}
\begin{frontmatter}
\title{Neutrinos and gamma-rays of hadronic origin from AGN jets} 
\author[UdM]{A. M. Atoyan,}
\author[NRL]{C. D. Dermer}
\address[UdM]{CRM, Universite de Montreal, Montreal H3C 3J7, Canada}
\address[NRL]{E. O. Hulburt Center for Space Research, Code 7653,
Naval Research Laboratory, Washington, DC 20375-5352 }

\begin{abstract}
We discuss the fluxes of high energy neutrinos and gamma-rays expected
from AGNs if hadrons can be effectively accelerated to ultra-high
energies by their relativistic jets, as currently believed.  Fluxes of
multi-TeV neutrinos detectable by $km$-scale detectors like {\it
IceCube} could be expected from powerful blazars where strong
accretion-disk radiation is present in the AGN cores. Gamma-ray fluxes
of hadronic origin can be important for flares in the compact jets of
these sources up to GeV energies, but they will be insignificant for
BL Lac objects. Production of UHE neutral beams composed of neutrons
and gamma-rays can drive straight collimated jets in the intergalactic
medium on multi-kpc scales, which could be resolved by the Chandra
X-ray observatory. While we do not expect any significant neutrino
flux from these large-scale jets, we predict gamma-rays of synchrotron
origin in the energy range from sub-GeV up to TeV energies, which
would be detectable by GLAST and ground-based gamma-ray telescopes.
\end{abstract}
\end{frontmatter}

\section{Introduction}

AGN jets, along with GRBs, are powerful
accelerators of relativistic particles, and the prime candidate
sources of extra-galactic cosmic rays (CRs) with energy spectra
extending beyond $10^{20}\,\rm eV$.  Detections of non-thermal flares
from blazars in the X-ray and particularly in the $\gamma$-ray domain
in the last decade have convincingly demonstrated that the compact
inner ({\it sub-parsec scale}) jets of blazars do accelerate particles
to very high energies \cite{har99,wee00}, presumably due to
relativistic shocks.  Although analyses of correlated X-ray and
TeV gamma-ray flares in BL Lacs support leptonic models
\cite{mk97,cat97,pia98,kca02}, which imply efficient acceleration of
relativistic electrons, the associated acceleration of hadrons is
expected with at least the same efficiency as that of the
leptons. It might be different in electron-positron pair jets where few
hadrons would be present, but comparison of the radio lobe and inner
jet powers indicates that jets are composed mainly of electrons and
protons \cite{cf93}, so that a nonthermal hadronic component is very
plausible.

Observational evidence for acceleration of hadrons in the jets of
AGN requires a target for interactions 
in the jets. In principle, ultrarelativistic protons accelerated in
the inner jets up to energies $\sim 10^{20}\,\rm eV$ could produce
synchrotron emission detectable at TeV energies \cite{aha00,mp01}, but
this requires extremely strong magnetic fields $\sim 20$-100 G in the
sub-parsec scale jets.  All other observable consequences of hadron
acceleration result from interactions with ambient material or
photon fields.  These include neutrinos, detection of which would
provide a direct proof for hadron acceleration, and the high
energy electromagnetic radiation from the secondary leptons and
$\gamma$-rays. Such evidence could also be provided by production of
collimated beams of ultra-high energy (UHE; $\gg
10^{15}\,\rm eV$ for the discussion below) neutrons and gamma-rays
formed in the same process of hadronic interactions, which could
explain effective transport of energy from AGN cores to multi-kpc scales
\cite{ad03}.

\section{Neutrinos and gamma-rays from compact jets}

Interaction cross sections of relativistic protons and nuclei with
 targets in astrophysical environments are generally not as large 
as cross sections of relativistic electrons. Therefore one would
 expect that observable radiation of hadronic origin might
be produced first of all in compact relativistic jets at sub-parsec
scales where the target, be it gas or photons, still remains
sufficiently dense.  One group of hadronic jet models invokes nuclear
interactions with ambient matter \cite{bb99,sps02} through the process
$p+p \rightarrow
\pi^{\pm, 0} \rightarrow\nu, e^{\pm}, \gamma$.
 Nuclear interaction models require, however, large masses, resulting
 in uncomfortably large total energy in the jet \cite{ad03}, which
 includes kinetic energy of the blob and the energy of relativistic
 protons needed for production of the observed $\gamma$-ray fluxes.

A second group of hadronic models is based upon photomeson
interactions of relativistic hadrons with ambient photon fields in the
jet. The relevant proton-photon processes are $ p + \gamma
\rightarrow p + \pi^{0}$ followed by $\pi^{0} \rightarrow
2\, \gamma$, and $p + \gamma \rightarrow n + \pi^{+}\;$, followed by
$\pi^{\pm} \rightarrow
\mu^{\pm} + \nu_{\mu} \rightarrow e^{\pm} + 2 \, 
\nu_{\mu} + \nu_{e}$. The same pool of secondaries is produced in multi-pion
production channels, and it is important to notice that 
in about half of all these inelastic collisions the primary relativistic
proton will be converted to a relativistic neutron.

Most of the photohadronic models take into account collisions of
high-energy protons with the internal synchrotron
photons \cite{mb92,man93,muc02}, while others also take into account
external radiation that originates either directly from the accretion
disk \cite{bp99} or from disk radiation that is scattered by
surrounding clouds to form a quasi-isotropic radiation
field \cite{ad01}. BL Lac objects have weak emission lines, so in these
sources the dominant soft photon field is thought to be the internal
synchrotron emission. The strong optical emission lines from the
illumination of BLR clouds in FSRQs reveal luminous accretion-disk and
scattered disk radiation.

The photomeson interaction threshold is about 150 MeV in the rest
frame of the proton, which requires ultra-high energies for the
accelerated particles to be above threshold for optical/UV photons.
In the case of internal synchrotron radiation, the energy output of
secondary particles formed in photohadronic processes is generally
peaked in the energy range from $\approx 10^{16}$-$10^{18}$ eV in
either low- or high-frequency peaked BL Lac objects \cite{muc02}, which
implies that such models can only be efficient if protons are
accelerated to even higher energies. This demand upon proton
acceleration for efficient photomeson production on the {\it internal}
synchrotron photons also holds for FSRQs, which have similar
nonthermal soft radiation spectra as low-frequency peaked BL Lac
objects.  However the presence of the isotropic external radiation
field of UV/soft X-ray energies in the central $\sim 0.1$-1 parsec
broad-line emission region (BLR) of FSRQs
 both substantially increases the photomeson
production efficiency and relaxes the requirement of very high proton
energies needed for $p\gamma$ interactions \cite{ad01}\,.
Another important feature of the external radiation target is that,
unlike the internal radiation, it does not dilute in
the course of rapid expansion of the jet/blob, and is available for
photomeson interactions during all the time of propagation
of UHE protons through the BLR.

In the model of Atoyan and Dermer \cite{ad01,ad03}, protons are 
accelerated in an outflowing plasma blob moving with bulk
Lorentz factor $\Gamma$ along the symmetry axis of the
accretion-disk/jet system. These  protons are assumed to
have an isotropic pitch-angle distribution in the comoving frame of a
plasma blob, within which is entrained a tangled magnetic field. Using
the parameters of the non-thermal flares detected from the FSRQ blazar
3C279 and from the BL Lac object Mkn 501 as prototypes, the fluxes of
high-energy neutrinos and gamma-rays (of hadronic origin) are
calculated \cite{ad03} assuming that the power to accelerate
relativistic protons is equal to the power injected into relativistic
electrons needed to explain the non-thermal flares detected by
EGRET \cite{weh98} and BeppoSax \cite{pia98}.

\begin{figure}
\centerline{
\epsfxsize=6.4cm \epsfbox{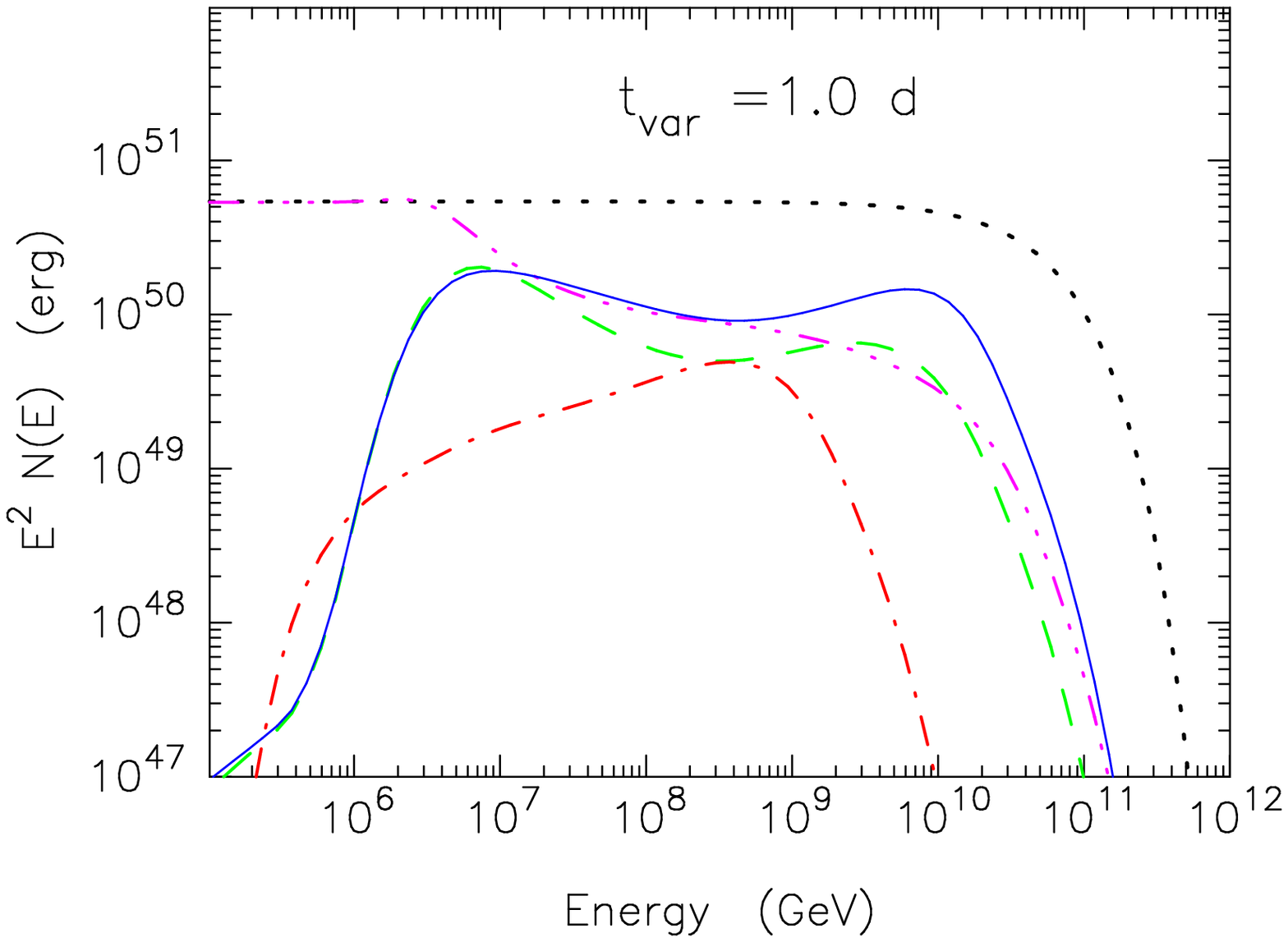}\hspace{5mm}
\epsfxsize=6.cm \epsfbox{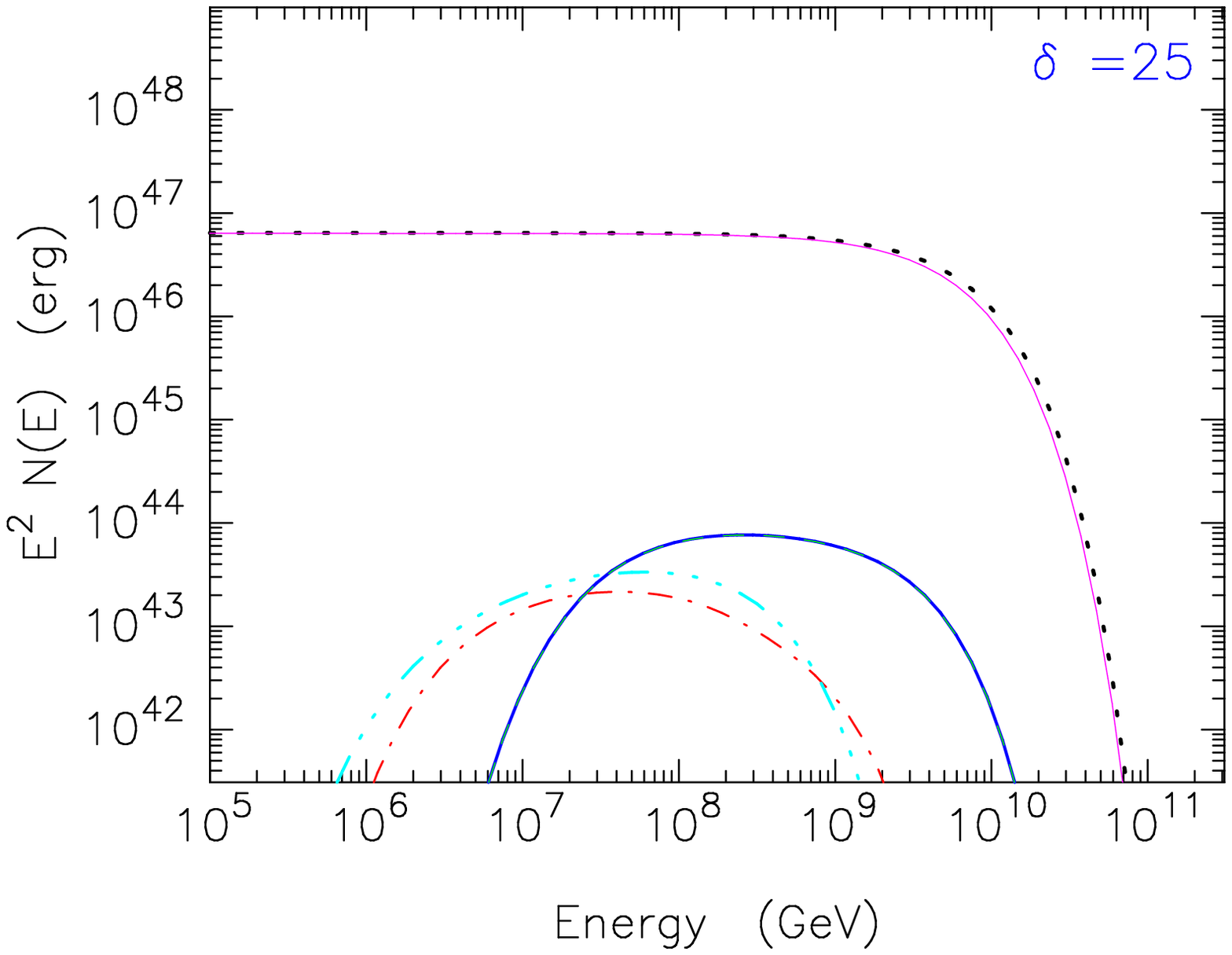}
}
\caption{
{\bf a }(left panel):~~ Spectra of protons injected into the blob
(dotted curve), protons which remain in the blob when it reaches the
edge of the BLR (thin solid curve), neutrons escaping from the blob
(thick solid curve), and escaping neutrons (dashed curve) and gamma
rays (dot-dashed curves) which reach the edge of the BLR, calculated
for 3C279 flare parameters assuming $\delta=10$; {\bf b }(right
panel):~~ Similar particle spectra as on the left panel, but
calculated for the parameters of the flare of Mkn 501, assuming
$t_{var} = 0.3 \,\rm d$ and $\delta = 25$. In addition, here we also show 
by 3-dot--dashed curve the spectra
of neutrinos produced in the blob.
}
\end{figure}

Fig.~1a ({\it left panel}) shows the spectra of neutrons escaping the
blob (heavy solid curve) and then the BLR (dashed curve) for 3C279,
calculated for the variability time $t_{var} = 1\,\rm d$ and the jet's
Doppler-factor $\delta = 10$, which defines the blob size $R\leq t_{var} c
\delta /(1+z)$. The dot-dashed curve shows the spectrum of the
gamma-rays produced by the neutrons outside the blob (but within the
BLR), and which therefore can escape the absorption on the internal
synchrotron photons inside the blob.  The dotted curves show the
overall spectra of protons injected into the blob, assuming that
injection has occurred during the time $ \simeq t_{var}$, and the thin
solid curves show the spectra of protons which remain in the blob by
the time that the blob reaches the edge of the BLR. The fraction of the
total energy of injected protons which is taken away by the UHE
neutrons with $E_n \geq 10^{17}\,\rm eV$ and the $\gamma$-rays with
$E_\gamma \geq 10^{16}\,\rm eV$ is generally at the level of $\sim 10
\%$ of the total energy injected in UHE protons. These numbers become
dramatically lower, dropping down by orders of magnitude, in the case of
BL Lac-type objects where the external radiation field is almost
absent, so that the intensity of photomeson interactions relies only
upon the internal synchrotron field. This is demonstrated in Fig.~1b
using flare parameters of Mkn~501. Note that the UHE gamma-ray fluxes
shown in Fig.~1b by dashed-dotted curve are produced predominantly
{\it inside} the blob, but because of much lower density of the
internal radiation these gamma-rays may have a chance
(depending also on $\delta$) to avoid $\gamma\gamma$ pair-production
absorption.

\begin{figure}[t]
\centerline{  \epsfxsize=7.3cm \epsfbox{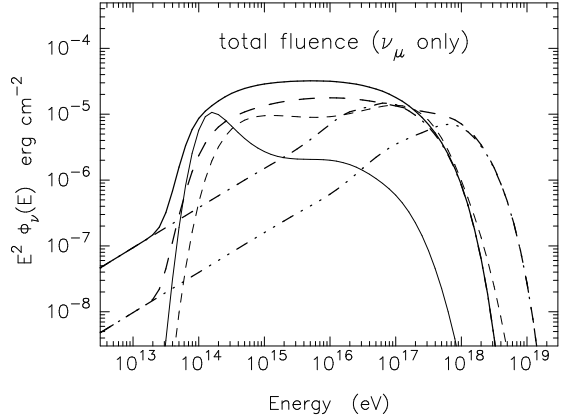}}   
\caption{Fluences of neutrinos integrated over several days
in the observer frame determined by the time for the blob to pass
through the BLR in 3C279. The solid and dashed curves show the fluences
calculated for $\delta = 6$ and 10, and the thick and thin curves
represent the fluences of neutrinos produced by photopion interactions
inside and outside the blob, respectively. The dot-dashed and 3-dot --
dashed curves show the fluences due to $p\gamma$ collisions if
external radiation field is not taken into account.}
\end{figure}

\begin{figure}[ht]
\centerline{\epsfxsize=7.3cm  \epsfbox{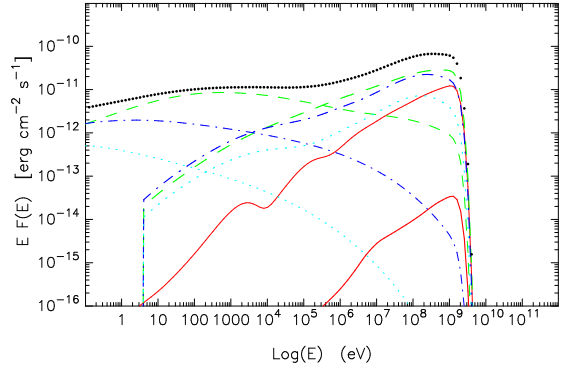}}
\caption{Radiation flux produced in and escaping from the blob (full dots) 
following the electromagnetic cascade initiated by energetic electrons
and gamma rays produced in photopion interactions, calculated for the
parameters of 3C 279 flare in case of $\delta = 10$.  The thick and
thin curves correspond to synchrotron and Compton radiations,
respectively. The radiation of the first generation of electrons,
which includes the
electrons produced by absorption of $\pi^0$-decay $\gamma$-rays in the
blob, is shown by the solid curves.  The dashed, dot-dashed and
3-dot--dashed curves show contributions from the 2nd, 3rd and 4th
generations of cascade electrons, respectively.}
\end{figure}

Fig.~2 shows the integrated neutrino fluences over the time it takes
for the blob to pass through the BLR in 3C 279.  The solid and dashed
curves show the fluence calculated for $\delta = 6$ and 10,
respectively. The thick and thin curves represent the fluence of
neutrinos produced in photomeson interactions inside and outside the
blob, respectively. For the spectral fluence shown in Fig.~2, the
total number of neutrinos that could be detected by a km$^3$
detector such as {\it IceCube}, using calculated neutrino detection
efficiencies \cite{ghs95}, are 0.37 for $\delta = 6$, and 0.21 for
$\delta = 10$. These numbers could be several times larger if the jet
power that goes to acceleration of protons is higher than that of the
electrons, implying hadronically dominated jets. We note, however,
that the neutrinos can be expected not only during the flares, but
also in periods between the active flares, insofar as the EGRET
observations show a persistent {\it average} level of gamma-ray fluxes
from 3C 279 at the level of $\sim 10$\% of the flare flux, implying
persistent electron/proton acceleration. Since the external photon
target is always available, over the course of one year several
neutrinos should be detected from FSRQs such as 3C 279 with km-scale
neutrino detectors.  The presence of a quasi-isotropic external
radiation field becomes crucial for neutrino detection, which results
in the prediction that FSRQs will be detected with km-scale neutrino
detectors, whereas BL Lac objects are much less promising for neutrino
detection \cite{ad01,ad03}.

We also calculated \cite{ad03} the associated radiation from the
cascades induced by photohadronic processes in the compact
jets. Fig.~3 shows the multiple generations of cascade radiation and
the total emergent flux associated with the neutrino production given
in Fig.~2\,. Purely hadronic models could be
effective for interpretation of gamma-ray fluxes only up to GeV energies
observed from FSRQs by EGRET. At higher energies, the hadronic
fluxes would be absorbed in the external UV
radiation field if it is sufficiently dense to be effective
for photomeson interactions. If the source is transparent
to TeV radiation, as in case of BL Lac objects, the intensity of
photohadronic processes would be consequently low.  Fig.~2 also
shows that the efficiency of proton acceleration is limited by the
intensity of the cascade radiation. In particular, X-ray observations
of 3C 279 limit the ratio of proton-to-electron acceleration powers
in the flaring state to a factor $\leq 10$.

\section{Radiation from the large-scale jets}

Because the UHE neutrons and gamma-rays are produced
quasi-isotropically in the comoving frame of a relativistically moving source
with $\Gamma_{jet} \sim 10$, for the external observer they are
collimated in a beam with a characteristic opening angle $\theta \sim
1/\Gamma \sim $ few degrees. As our calculations show \cite{ad03}, the
energy contained in this neutral beam in case
of FSRQs can reach a few percent of the total energy injected in accelerated
hadrons in the compact sub-parsec jet, which is very
substantial. Thus, for the jets in 3C 279 this would
correspond to a mean power for UHE neutral beam production
$\sim 10^{45} \,\rm erg/s$. This power is then transported to
large distances from the compact nucleus, and is released in the
processes of neutron decay \cite{ad01,ad03}, $n \rightarrow p + e +
\nu$, and $\gamma\gamma\rightarrow e^\pm$ attenuation \cite{ner02,ad01}
on the cosmic microwave background radiation field. The neutron decay
paths are $l_{n.d.} \simeq 1 \, (E/10^{17} \,\rm eV) \,\rm kpc$,
i.e. in the range $\sim {1\,\rm kpc} - 1\,\rm Mpc$ for $E_{n} \sim
10^{17}$-$10^{20}\,\rm eV$.  This is similar to the $\gamma
\,\gamma$-interaction pathlengths for the gamma-rays with energies
$10^{16}$-$10^{19}\,$eV. These decay distances agree with the length
scales of the straight narrow jets resolved by the Chandra X-ray
observatory from a number of AGN. Therefore neutral beams may
provide a new interpretation for the origin of these jets, given
that there is a mechanism for the conversion of the energy and
momentum of the charged particles emerging from the neutral beam into
the surrounding intergalactic medium, with subsequent
electron acceleration out of the thermal pool. The latter is
required for the interpretation of the synchrotron fluxes of these
jets detected at radio and optical wavelengths.

For the UHE gamma-ray component of the neutral beam, we note that the
 emerging $e^+ - e^{-}$ pairs would almost immediately lose their
 energy by synchrotron radiation in the ambient magnetic fields with a
 transverse (to the jet direction) component $B_{\perp} \sim  1\,\rm
 \mu G$ or larger. The cooling path of these electrons, $l_{cool}$, is
 typically much shorter than their gyroradius: $$l_{cool}/R_{gyr}
 \simeq 0.2 (B_{\perp}1\,\rm \mu G)^{-1} \, (E_{e}/10^{17}\,\rm
 eV)^{-2}\,,$$ 
which implies that these electrons would not exert any
 significant pressure on the ambient medium. The energy of these
 electrons will be converted into $\gamma$-rays with
 characteristic energies $ \epsilon_{syn} \simeq 4
 \,(B_{\perp}/10\,{\rm \mu G}) \, (E_{e}/10^{17}\,\rm eV)^{2}\,
\rm GeV\, .$

 For the beam of parent gamma-rays with $E_\gamma $ from $ \sim
  10^{16}$ to $> 10^{18}\,\rm eV$, the synchrotron $\gamma$-ray
  fluxes in the characteristic jet magnetic fields  $ \sim 10 \,\rm \mu
  G $ should appear at sub-GeV to TeV energies. These synchrotron
 gamma-rays would be strongly polarized, the detection of which
  could support their
  hadronic origin.  Another, albeit less
  conclusive, feature of the synchrotron gamma-ray fluxes would
  consist in the detection of atypically hard spectral indices, with most
  of the energy flux peaked at high energies due to
  the sharp cutoff in the production spectra of $e^{+}-e^{-}$ pairs
  below $10^{16}\,\rm eV$.  These fluxes could be detectable by 
  GLAST at GeV energies, and at higher energies by the ground-based
  gamma-ray telescopes with energy thresholds well below 100 GeV, like
  HESS, VERITAS, etc.  The sensitivity of these instruments is at the
  energy flux level $f_{-12} \equiv f_{\epsilon}
\sim 10^{-12} \,\rm erg \, cm^{-2} \, s^{-2} \sim 1 $ or better 
\cite{wee00,iact}. This implies an apparent ``4$\pi$''
 luminosity of the source at a distance
$d = d_{Gpc} \,\rm Gpc$ at the level of $L_{4\pi} \simeq 10^{44}
f_{-12} \, d_{Gpc}^2\, \rm erg/s$.  For comparison, the {\it absolute}
power of the neutral beam for the parameters used for 3C279 is at the
level $L_{beam} \sim 10^{45}\,\rm erg/s$.

 The energy and momentum of the neutron component of the beam after
the $\beta$-decay goes primarily to the protons. There is no effective
target for interactions of these protons with the intergalactic
medium, except via the magnetic field. Any significant deflection of
the protons from their initial direction implies that a beam of
$\beta$-decay protons should transfer its momentum to the surrounding
medium to drive turbulence which could effectively deposit both the beam
energy and momentum along its path. Driving the intergalactic medium into
motion will cause stretching of the ambient magnetic field, whatever
 its initial orientation, along the beam, as is observed in
large-scale jets.  Thus, the UHE neutrons of the neutral beam
represent the main component for driving and continuously supplying
power into the large scale jets.  Electrons from the ambient medium
accelerated in the first or second order Fermi process could explain
the synchrotron fluxes extending up to the X-ray
domain \cite{da02}. Interestingly, the $e^{+}\, e^{-}$ pairs from the
gamma-ray beam, or even the $\beta$-decay electrons of the neutron beam,
might lead to a two-component synchrotron models which seem to be
needed for consistent interpretation of optical/X-ray data in some
jets like Pictor A. 

It is important to note that this method of energy transport up to Mpc
scales by neutral UHE beam avoids quenching of the jet by material
both in dense central region on parsec scales, and in the
intergalactic medium on multi-kpc scales.  Acceleration of hadrons in
the compact jets of AGN also implies that most of the UHE protons
would accumulate in time in the central region of AGN at the parsec
scales after deceleration of the inner jet. A partial degradation of
the energy of these protons with $E_{p} \sim 10^{15} - 10^{16}\,{\rm eV}$
in the photomeson collision with the accretion disk radiation would
initiate Mpc-scale halos \cite{acv94} around such FSRQs, due to pair
cascades on the CMBR, which might be detectable by forthcoming
gamma-ray telescopes.

{\it Acknowledgments.} AA appreciates the financial support provided by the LOC. The work of CD is
supported by the Office of Naval Research and NASA grant DPR S-13756G.

\end{document}